\documentclass[aip,reprint]{revtex4-1}
\usepackage[T1]{fontenc}
\usepackage[utf8]{inputenc}
\usepackage{amsmath}
\usepackage{amssymb}
\usepackage{graphicx}
\usepackage{multirow}
\usepackage{soul}
\usepackage{color}
\usepackage[
citecolor=blue,
colorlinks,
linkcolor=blue,
urlcolor=blue]{hyperref}
\usepackage{epstopdf}
\usepackage{bm}
\usepackage{eqnarray}
\usepackage{subcaption}
\usepackage{graphicx}
\usepackage{xcolor}
\usepackage[font=small,labelfont=bf,tableposition=top]{caption}

\DeclareCaptionLabelFormat{andtable}{#1~#2  \&  \tablename~\thetable}
\draft 

\begin{document}

\title{Hypochaos prevents tragedy of the commons in discrete-time eco-evolutionary game dynamics}

\author{Samrat Sohel Mondal}
\email[]{samrat@iitk.ac.in}
\affiliation{Department of Physics, Indian Institute of Technology Kanpur, Uttar Pradesh 208016, India}

\author{Avishuman Ray}
\email[]{avishuma@usc.edu}
\affiliation{Department of Physics and Astronomy, University of Southern California, Los Angeles, California 90089, USA}

\author{Sagar Chakraborty}
\email[]{sagarc@iitk.ac.in}
\affiliation{Department of Physics, Indian Institute of Technology Kanpur, Uttar Pradesh 208016, India}

\newcommand{\coloredsout}[2]{\textcolor{#1}{\sout{#2}}}
\newcommand{\coloredstrikethrough}[2]{%
  \textcolor{#1}{\sout{\textcolor{black}{#2}}}%
}

\date{\today}
\begin{abstract}
	While quite a few recent papers have explored game-resource feedback using the framework of evolutionary game theory, almost all the studies are confined to using time-continuous dynamical equations. Moreover, in such literature, the effect of ubiquitous chaos in the resulting eco-evolutionary dynamics is rather missing. Here, we present a deterministic eco-evolutionary discrete-time dynamics in generation-wise non-overlapping population of two types of harvesters---one harvesting at a faster rate than the other---consuming a self-renewing resource capable of showing chaotic dynamics. In the light of our finding that sometimes chaos is confined exclusively to either the dynamics of the resource or that of the consumer fractions, an interesting scenario is realized: The resource state can keep oscillating chaotically, and hence, it does not vanish to result in the tragedy of the commons---extinction of the resource due to selfish indiscriminate exploitation---and yet the consumer population, whose dynamics depends directly on the state of the resource, may end up being composed exclusively of defectors, i.e., high harvesters. This appears non-intuitive because it is well known that prevention of tragedy of the commons usually requires substantial cooperation to be present.
\end{abstract}

\maketitle 
\begin{quotation}
When a population of social, economic, biological, or ecological interacting agents exploit a common resource selfishly, the resource is destroyed; one says that the tragedy of the commons has occurred. The depleted resource, in turn, naturally has adverse effect on the agents' survival. Hence, a feedback between the agents and the resources is set up. While the dynamics of the interaction between the agents can be understood using the game theory, the state of the resource can be modelled using using some standard population growth model, like the famous logistic equation. Thus, game-resource feedback dynamics is what one should investigate mathematically to understand the aforementioned scenario. In this paper, we specifically look into a population of agents and resource that are generation-wise non-overlapping---parents and offsprings do not exist together in any generation. This leads to a time-discrete coupled dynamical equations. Consequently, the simple deterministic dynamics is very rich with occurrence of both convergent and oscillatory---periodic and chaotic---outcomes. We present the intriguing interplay between the chaos and the tragedy of the commons. The main goal is to show that chaos can help prevent the tragedy of the commons even in the complete absence of cooperators.
\end{quotation}
\section{Introduction}
\label{sec:introduction}
The world we live in is full of situations where groups, communities or populations  compete for a shared resource in ecological, socio-economical, or political scenarios. However, conservation of a shared resource is in direct conflict with private interest of an individual. This conflict leads to the recurring theme of the tragedy of the commons (TOC)~\cite{malthus,lloyd1833two,hardin1968s,ostrom1999coping}. The TOC occurs when individuals, interested solely on their personal gains, consume the shared common resource without any regard for the needs of larger community. This results in the inevitable destruction of the shared resource, leading to severe consequences for the community (which includes the individual) as a whole. Uncontrolled population growth~\cite{hardin1968s}, water pollution and crisis~\cite{Shiklomanov2000}, the contamination of earth's atmosphere~\cite{jacobson}, property and communal rights or state regulation~\cite{ostrom1999coping}, and wildlife crimes~\cite{Pires2011} are a few of the illustrating examples of the TOC. A more contemporary example could be seen during the COVID-19 outbreak when some persons avoiding vaccine shots (due to fear of rare side-effects and deciding to rely on herd immunity) or hiding infection (in order to escape quarantine), lead to the more severe spread of the disease~\cite{Maaravi2021}. The phenomenon of TOC is not restricted to the cognitively superior human population only---it can be widely witnessed in the biological world from {microbes\cite{Schuster2017,Smith2019}} to mammals~\cite{Rankin2006}.

It is obvious that the state of shared resource---i.e., how replete or deplete it is---can affect the evolutionary fitnesses of the types of individuals present in the population, consequently changing the preferences---cooperation or defection---of the individuals (henceforth, called players following the game-theoretic terminology). As the environment deteriorates, the cooperation tendency should develop in the population in order to avert the TOC. 

Several recent investigations~\cite{Melbinger2010,Cremer2011,Wienand2015,Chowdhury2021,NagChowdhury2021,Roy2022,Roy2023,NagChowdhury2023} have delved into the evolutionary dynamics of populations, specifically focusing on the ecological expansion of population size. However, a limited number of noteworthy studies~\cite{Wienand2017,Wienand2018,Becker2018} posit that the potential of such ecological expansions may be contingent on available resources. These studies have explored the resource dependence of carrying capacity of ecological growth dynamics in conjunction with evolutionary processes. Within the paradigm of nonlinear dynamics, recent studies~\cite{weitz2016pnas,Chen2018,lin2019prl,tilman2020nc,Bairagya2021,Yan2021,Mondal2022,Liu2023,Bairagya2023} have explicitly mathematized this coupled dynamics of ecological resource and its evolving consumers to gain insights about the TOC.  Such dynamical models are aptly called the eco-evolutionary dynamics. In passing, it is worth directing readers to the reviews~\cite{Xia2023,Perc2013} that discuss works related to evolutionary dynamics of public goods---an important precursor to eco-evolutionary game dynamics.

In this paper, we aim to fill a lacuna in the aforementioned set of papers: We want to present a study on deterministic eco-evolutionary dynamics in hitherto overlooked generation-wise non-overlapping populations, i.e., rather than working with the time-continuous models studied till now in the literature, we plan to investigate the time-discrete deterministic eco-evolutionary dynamics. Non-overlapping populations are, of course, not as common but examples are not hard to find: Populations of certain plants~\cite{FernndezMarn2014}, insects~\cite{Godfray1989}, parasites~\cite{May1984}, and rodents~\cite{Boonstra1989} may be well-approximated to be of non-overlapping type. An intriguing technical aspect that discrete dynamical equations (henceforth, sometimes called maps) renders is the possibility of chaotic outcomes even in the scenario where the consumer population with only two types harvest a single resource. Thus, chaos induced prevention of TOC may be seen in such cases. 

In this context, we would like to bring a terminology of relevance, coined and used in this paper, to the readers' attention. We know about hyperchaos~\cite{Rossler1979,Matsumoto1986,Baier1990} which essentially means that the underlying dynamics is such that two of its largest Lyapunov exponents are positive. Thus, the Lyapunov dimension, an estimate of the capacity dimension as per the Kaplan--Yorke conjecture~\cite{Kaplan1983}, of a hyperchaotic attractor must be greater than two. In fact, if all the Lyapunov exponents of an $N$-dimensional map are arranged as a finite sequence, $\{\lambda_k\}_{k=1}^{N}$, in descending order of magnitude, then the largest index $k=K$, for which $\sum_{k=1}^{K}\lambda_k\ge0$, is conjectured to be a lower bound of the attractor's capacity dimension. Can there be maps where the chaotic attractor's capacity dimension is lower than this lower bound? Apparently, the answer is in affirmative as we shall witness in this paper; such chaotic attractor may appositely be said to be have arisen from \textit{hypo}chaos.

Without further ado, let us delve into the precise mathematical description of the model which forms the backbone of our study in this paper.

\section{Construction of the model}
\label{sec:model}

Let there be only two distinct strategies that can be adopted by any individual in a consumer population with non-overlapping generations. Size of the population $N$ is considered to be constant, and practically infinite, throughout all time $t$. The population is furthermore considered to be well-mixed and unstructured. As the model considers non-overlapping generations, time $t$ is also considered to be discrete: $t\in\mathbb{N}$. Thus, the consumer population that adopts $i$th strategy at time $t$ is denoted by $N_{i}^t$ and hence, the frequency of $i$th strategy being used may be defined as $x_{i}^t\equiv N_{i}^t/N$, where $i \in \{1,2\}$. At any time instant, the row vector $\boldsymbol{x}^t\equiv [x_{1}^t,\,x_{2}^t]$ represents the instantaneous state of the consumers.

Next, let $m^t$, a non-negative real number, denote the state of the shared resource being consumed at time $t$. The composite system of the consumer population and the shared resource is, thus, given by $\boldsymbol{s}^t\equiv[\boldsymbol{x}^t,\,m^t]$. Consequently the model inherits three variables, viz., $x_{1}^t$, $x_{2}^t$, and $m^t$. However, since $x_{1}^t$ and $x_{2}^t(=1-x_{1}^t)$ are not independent variables, we henceforth use $x^t(=x_1^t)$ as the only variable to represent the state of the consumer population $\boldsymbol{x}^t$.

In the simplest nontrivial description, the unfortunate rise of the defector type of the consumers who consume the shared resource at a greater rate than the cooperators and lead the system to the TOC, can be exemplified through the Prisoner's Dilemma game~\cite{1965_RC}. The Prisoner's Dilemma game is a one-shot two-player--two-strategies game in which the strategy `defect' is the only symmetric Nash equilibrium~\cite{nash1950pnas,wagner2013} which turns out to be a non-Pareto-optimal one~\cite{1896_Pareto}. The corresponding payoff matrix can be written as
\begin{eqnarray*}
	\label{payoffTable} 
	\centering
	\begin{tabular}{cc|c|c|}
		& \multicolumn{1}{c}{} & \multicolumn{2}{c}{{Player $2$}}\\
		& \multicolumn{1}{c}{} & \multicolumn{1}{c}{Cooperate} & \multicolumn{1}{c}{\,\,\,\, Defect \,\,\,\,}\\\cline{3-4} 
		\multirow{2}*{{Player $1$}} & Cooperate & $R,R$ & $S,T$ \\\cline{3-4}
		& Defect & $T,S$ & $P,P$ \\\cline{3-4} 
	\end{tabular}\quad
\end{eqnarray*}
where the first and the second entries in each cell are the payoffs of player $1$ and player $2$ respectively. $R$, $S$, $T$, and $P$ respectively refer to Reward, Sucker's payoff, Temptation, and Punishment. In line with our model, we identify $x_{1}^t$ and $x_{2}^t$ as the instantaneous fractions of cooperators and defectors respectively. 
\subsection{Dynamical Equations}
\label{sec:formulation}
\subsubsection{Time evolution of resource}
In order to examine the fate of the shared resource, we need to consider the dynamics of resource harvesting. A considerable amount of studies discussed about one-dimensional logistic harvesting models~\cite{Cooke1986,Murray1993}. Of course, there are more realistic and improved models like the Ricker model~\cite{ricker1963handbook}, the Beverton--Holt model~\cite{beverton1957dynamics} and the Hassell model~\cite{Hassell1975,Hassell1976}; however, for the purpose of the present work, sticking with the logistic model is sufficient.

Our model assumes that the dynamics of the shared resource is governed by two factors, viz., there is an intrinsic growth rate $r$ of the shared resource by virtue of which it tends to grow `logistically'~\cite{May1984} up to its carrying capacity $k$, and a negative inhibitory feedback due to the harvesting of the resource by consumers. We use a time-discrete version of the equation used in a recent important study~\cite{tilman2020nc} to model the aforementioned resource dynamics, which is as follows:
\begin{equation}
\label{resourceDynamics}
\begin{split}
m^{t+1} =&~~rm^t\left(1-\frac{m^t}{k}\right)- m^t\left[e_{L}x^t + e_{H}\left(1-x^t\right)\right],
\end{split}
\end{equation}
where the resource state can take only non-negative real values, and so can $r$, $e_L$ and $e_H$. Here, $e_{L}$ $(e_{H})$ denotes the harvesting efforts of the cooperators (defectors). As, by definition, the defectors consume the resource at a greater rate than the cooperators, we have $e_{H}>e_{L}$; furthermore, for simplicity, all the parameters are considered to be independent of time. 

The harvesting term in the equation above may be understood as follows: It is not quite physical that all the consumers together harvest the finite-sized shared resource at any time instant; it is more reasonable to assume that at every time step, the resource is harvested by a finite random fraction of the entire consumer population. Due to the inherent assumption of a well-mixed unstructured population, the fractions of types of consumers harvesting at any instant are present in exactly the same proportions as in the entire population. Consequently, the consumers of two types---low and high harvesters---must be depleting the resource at a rate, $xe_L + (1-x)e_H$, dependent only on the fractions of consumer-types.

We obviously should restrict ourselves to the cases where non-negative values of $m$ at any $t$ are not allowed. The detailed analysis presented below reveals that we can identify a transformation $m^t\to n^t\equiv m^t/ \left[k(1-e_H/r)\right]$ so that the resultant map, viz.,
\begin{equation}
\label{resourceScaled}
\begin{split}
n^{t+1} =& ~rn^t\left[1-\left(1-\frac{e_H}{r}\right)n^t\right]-n^t\left[e_Lx^t+e_H\left(1-x^t\right)\right],
\end{split}
\end{equation}
renders the interval $[0,1]$ forward-invariant, provided 
\begin{subequations}
\label{conds}
\begin{eqnarray}
&&e_H\in \left(0,\infty \right),\label{ehrange}\\
&&e_L\in \left[e_H-1,e_H\right),\label{elrange}\\
{\rm and}\,\,&&r\in\left(r_-,r_+\right);\label{rrange}
\end{eqnarray}
\end{subequations}
where
\begin{equation}
	r_\mp \equiv (e_L+2)\mp 2\sqrt{\left(e_L+1\right)-e_H}\,.\label{eq:rmp}
\end{equation}
One notes that $r_{-}$ is always larger than $e_H$. Henceforth, we shall be concerned only with Eq.~(\ref{resourceScaled}) as far as the resource's dynamics is concerned. 
\subsubsection{Forward invariance of resource state}
\label{forwardResource}
\begin{figure}
	\centering
	\begin{subfigure}{0.8\linewidth}
		\includegraphics[width=\linewidth]{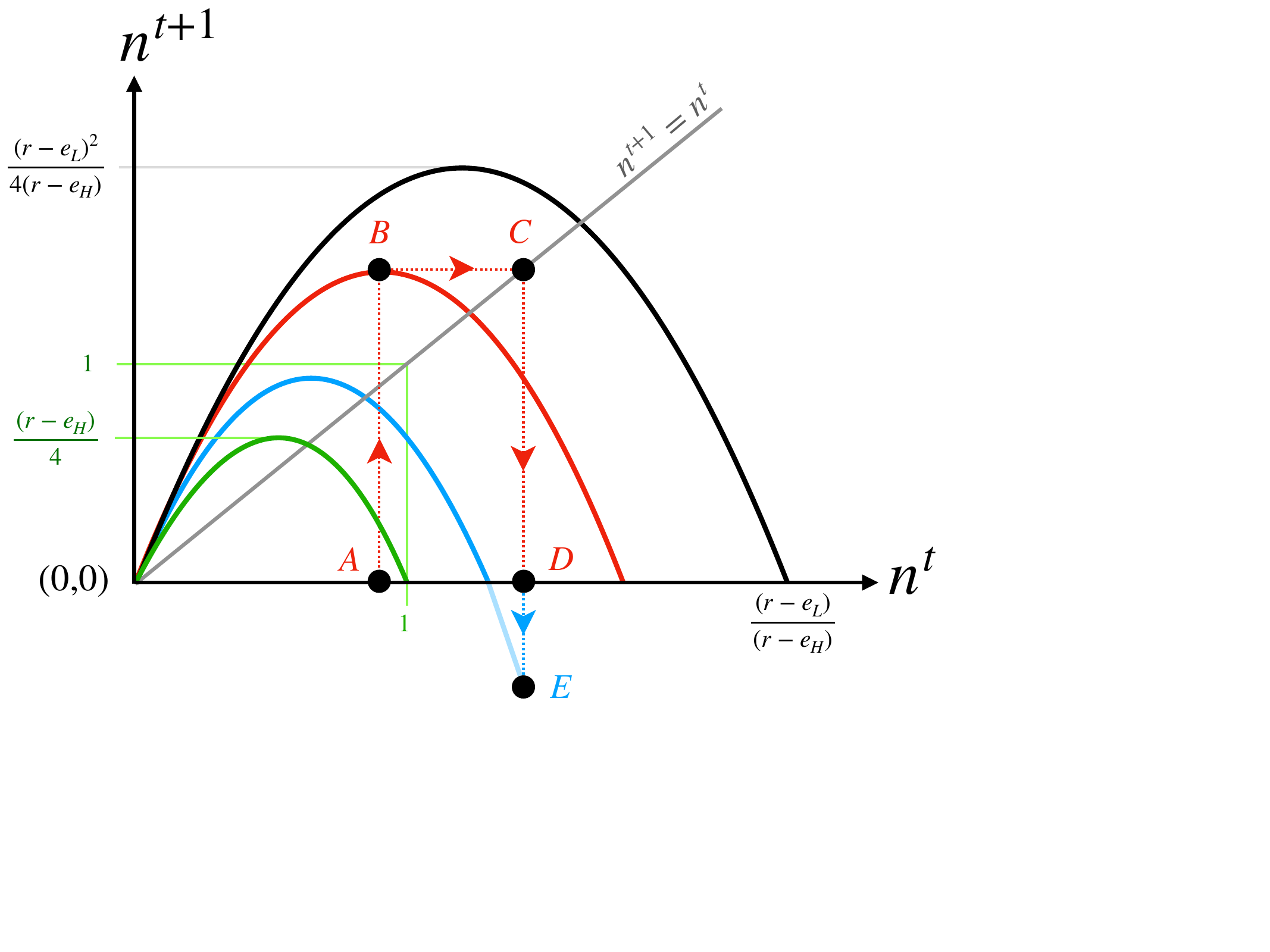}
		\caption{The case where inequality~(\ref{eq:11}) is not satisfied.}
		\label{fig:6a}
	\end{subfigure}
	\begin{subfigure}{0.8\linewidth}
		\includegraphics[width=\linewidth]{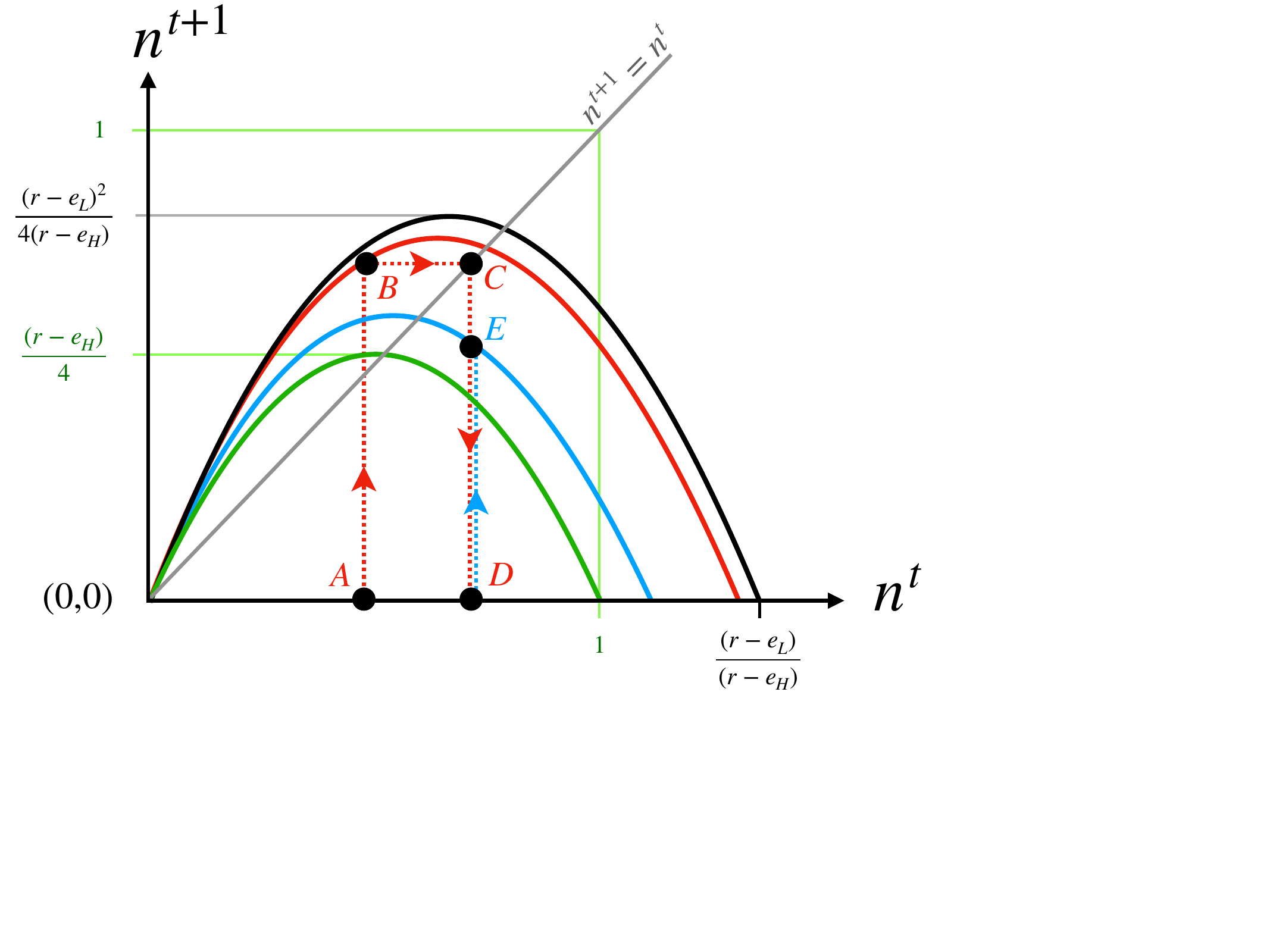}
		\caption{The case where inequality~(\ref{eq:11}) is satisfied.}
		\label{fig:6b}
	\end{subfigure}
	\caption{Cobweb diagrams for the resource dynamics geometrically interpreting inequality~(\ref{eq:11}). The green and black parabolas, respectively, are the parabolas with the minimum and the maximum attainable ordinate-values as per Eq.~(\ref{resourceScaled}). The base of the green parabola has a unit length. Assume that the value of $x^t$ at a time step $t$ is $x_{\rm red}$ (say) and the corresponding parabola for $n^{t}$ is the red coloured one; $n^{t}$ at $t$ is denoted by point $A$. At the subsequent time step, $t+1$, $x^{t+1}$ is $x_{\rm cyan}<x_{\rm red}$ (say); the corresponding parabola for $n^{t}$ is cyan coloured and $n^{t}$ at $t+1$ is denoted by point $E$---reached as a trajectory, ABCDE, in the cobweb diagram. See Sec.~\ref{forwardResource} for further details. Subplot (a) depicts that when inequality~(\ref{eq:11}) is not satisfied, indicating that the maximum of the black curve does not fall within the unit (green) square, iteration of resource state from point $D$ results in unphysical (negative) value for the resource (observe point E). Whereas, subplot (b), illustrates that when the maximum of the black curve lies completely inside the unit (green) square, the values of $n^{t}$ remain physically valid for all $x^{t}$ at all times.}
	\label{fig:6}
\end{figure}
In order to identify the ranges of the parameters such that the forward invariant interval of the (properly normalized) resource state evolving under Eq.~(\ref{resourceDynamics}) $\forall x^t\in[0,1]$ is $[0,1]$,  let us first define $a^t(x^t)\equiv e_Lx^t+e_H\left(1-x^t\right)$ which measures the rate of feedback from the consumer population. Consequently, we can recast Eq.~(\ref{resourceDynamics}) as 
\begin{equation}
\label{eq:8}
\begin{split}
m^{t+1} = \left[r-a^t\left(x^t\right)\right]m^t \left[1-\frac{m^t}{k\left(1-\frac{a^t\left(x^t\right)}{r}\right)}\right].
\end{split}
\end{equation}
Obviously, it has the form of a logistic map with a time-dependent modified carrying capacity, $k^t\equiv k\left(1-a^t\left(x^t\right)/r\right)$ and a time-dependent intrinsic growth rate, $r^t\equiv\left(r-a^t\left(x^t\right)\right)$. 

To ensure the non-negativity of resource state $m^t$, it is imperative that the modified intrinsic growth rate $r^t$ remains non-negative at all times (i.e., for values of $x^t$). Mathematically,
\begin{eqnarray}
&&\min_{\substack{0 \leq x^t \leq 1 }} r^t(x^t) \geq 0, \nonumber\\
\implies &&r \geq e_H.\label{eq:10}
\end{eqnarray}

Since the rate of change of $k^t$ with respect to $x^t$ is $k(e_H-e_L)/r$, we can conclude that $k^t$ is a strictly increasing function of $x^t$. Consequently, the minimum value of $k^t$ is $k_{\mathrm{min}}\equiv k(1-e_H/r)$. From the term, $[1-m^t/k^t]$, in Eq.~(\ref{eq:8}), it is clear that $m^t=k_{\mathrm{min}}$ is largest allowed value for which $m^{t+1}$ remains non-negative for any value of $x^t$. Therefore, the transformation $m^t\to n^t=m^t/k_{\mathrm{min}}$ scales interval $[0,k_{\min}]$ to fill the range: $[0,1]$. The modified equation for resource dynamics now reads
\begin{equation}
\label{eq:9}
n^{t+1}=\left[r-a^t(x^t)\right]n^t\left[1-\frac{n^t}{\frac{\left(r-a^t(x^t)\right)}{\left(r-e_H\right)}}\right].
\end{equation}
Clearly, it has the form of a logistic map with a time-dependent modified carrying capacity, $\tilde{k}^{t}(x^t) \equiv \left(r-a^t\left(x^t\right)\right)/\left(r-e_H\right)$ and a time-dependent intrinsic growth rate, $r^t$.

In order that interval $[0,1]$ is forward invariant under the action of map (\ref{eq:9}), it is required (see Fig.~\ref{fig:6}) that
\begin{eqnarray}
&&\max_{\substack{0 \leq x^t \leq 1 }}\left[ \max_{\substack{0 \leq n^t \leq 1 }} n^{t+1}\right] \leq \min_{\substack{0 \leq x^t \leq 1 }} \tilde{k}^{t}(x^t)\nonumber\\
\implies &&\frac{\left(r-e_L\right)^2}{4\left(r-e_H\right)} \leq 1.\label{eq:11}
\end{eqnarray}
To comprehend this, note that for a fixed $x^t$, $n^t$-$n^{t+1}$ plot is a parabola, such that the maximum of the parabola, ${\left[r-a^t\left(x^t\right)\right]^2}/{[4(r-e_H)]}$,  and the rightmost point of the parabola, $\tilde{k}^t$ increase with increase in the fixed value of $x^t$. In fact, the biggest-sized parabola (see the black curve in Fig.~\ref{fig:6}) has its maximum with ordinate-value ${(r-e_L)^2}/{[4(r-e_H)]}$ and it crosses abscissa at $(r-e_L)/{(r-e_H)}$ [see inequality~(\ref{eq:11})]. Similarly, the smallest-sized parabola (see the green curve in Fig.~\ref{fig:6}) has maximum ordinate-value ${(r-e_H)}/4$ and it crosses abscissa at $1$. Therefore the geometrical meaning of inequality~(\ref{eq:11}) is essentially that the ordinate-value of the maximum of the black parabola can not be greater than the maximum abscissa-value of the green parabola. 

Fig.~\ref{fig:6a} succinctly exhibits the problem in case the inequality is not satisfied: Suppose, as shown in the figure, the black curve lies outside the green curve. Then for some values of $x^t$, there can be parabolae that lie within these two curves. For illustration, we consider the red and the cyan-colored parabolae; the former is for $x$-value ($x_{\rm red}$, say) that is greater than that ($x_{\rm cyan}$, say) of the latter. A phase point ($x_{\rm red}$, $n^t$)---where $n^t\in[0,1]$---at time $t$, denoted by $A$ in Fig.~\ref{fig:6a} should be mapped to point $D$ at time $t+1$ via $B$ and $C$. If at $t+1$, the co-evolving $x^t$ changes its value to $x_{\rm cyan}$, then on the next iteration $n^t$ must take a negative value as depicted by point $E$---an unphysical behavior. The source of this unphysical behaviour is straightforward: $n^{t+1}$ can not be greater than the maximum of the parabola (given by Eq.~(\ref{eq:11}) for a fixed $x$-value) used to evolve the corresponding initial $n^t$. Fig.~\ref{fig:6b} illustrates the case where inequality~(\ref{eq:11}) is satisfied: Since all the parabolae must lie between the green and the black ones, every $n^t$ point in the range $[0,1]$ must be mapped back into the same range.

Finally, solving inequalities~(\ref{eq:10}) and (\ref{eq:11}) yields a range $(r_{-},r_{+})$, where $r_{\mp}$ is given in Eq.~(\ref{eq:rmp}), that renders $[0,1]$ forward invariant. Since the intrinsic growth rate, $r$, can not be complex-valued, $e_L$ must not be less than $(e_H-1)$. In effect, we have derived conditions~(\ref{conds}).

\subsubsection{Resource-dependent replicator equation}
Next, in order to get to the replicator equation~\cite{} corresponding to the high and low harvesters---being synonymously called cooperator and defector respectively in this paper---we start with the simplest situation where $R$, $S$, $T$, and $P$ are the payoffs realized ignoring the state of the environmental resource, $n^t$. In this case the payoff matrix, ${\sf U}(n^t)$, for a focal player would have the $n^t$-independent form:
\begin{eqnarray}
{\sf U}({n^t}={\rm constant})={\sf U}=
\left[ {\begin{array}{cc}
	R & S \\
	T & P \\
	\end{array} } \right]. 
\label{eq-U}
\end{eqnarray}
Subsequently, we extend the payoff matrix ${\sf U}$ to include the effect of state of environment~\cite{weitz2016pnas,tilman2020nc,Bairagya2021,Rankin2007} as follows:
\begin{eqnarray}
\label{effectivePaym}
{\sf U}^t(n^t) = (1-n^t) {\sf U_0} + n^t {\sf U_1}.  
\end{eqnarray} 
Here ${\sf U_k}$  is the shorthand notation for ${\sf U}(n^t=k)$ where $k\in\{0,1\}$. It is evident that the payoff matrix reduces to ${\sf U}_{0}$ in the limit of $n^t=0$, which corresponds to the poorest resource state. Whereas in the opposite limit of $n^t=1$ of the richest resource state, it reduces to ${\sf U}_{1}$. Guided by the form of Eq.~(\ref{eq-U}), a natural parametrization of ${\sf U}_{0}$ and 
${\sf U}_{1}$ is 
\begin{eqnarray}
{\sf U}_{0}=
\left[ {\begin{array}{cc}
	R_0 & S_0 \\
	T_0 & P_0 \\
	\end{array} } \right]~
\label{eq-U0}
\end{eqnarray}
and 
\begin{eqnarray}
{\sf U}_{1}=
\left[ {\begin{array}{cc}
	R_1 & S_1 \\
	T_1 & P_1\\
	\end{array} } \right].
\label{eq-U1}
\end{eqnarray}

We note that the matrix ${\sf U}^t$ is independent of $x^t$. However, our model considers matrix games~\cite{cressman2014pnas}, for which $x^t$-dependence enters through the fitness. The fitness $f_{i}^t$ of an individual consumer adopting the $i$-th strategy is given by
\begin{equation}
f_{i}^t(x^t) = \sum_{j=1}^{2} {\sf U}_{ij}^t({ n^t}) x_{j}^t;\quad i=1,2.
\end{equation}
As the consumer population grows and evolves through a replication-selection process~\cite{2021MCChaos}, therefore it is appropriate to use the paradigmatic replicator maps to model its evolution. We use both type-I and type-II replicator maps~\cite{Mukhopadhyay2020, Pandit2018} which are

\begin{subequations}
	\label{replicator}
	\begin{eqnarray}
	\label{replicatorI}
	x^{t+1} &=& x^t + x^{t}\left(1-x^{t}\right)\left(f_{1}^t-f_{2}^t\right)\\
	\label{replicatorII}
		{\rm and}\,\,
x^{t+1}&=&x^{t}\frac{f_{1}^t}{\bar{f}^t},
	\end{eqnarray}
\end{subequations}
respectively. Here $\bar{f}^t\equiv\sum_{i=1}^2\sum_{j=1}^2 {\sf U}_{ij}^t(n^t)x_{i}^tx_{j}^t$ denotes the mean fitness of the population. 

Either Eqs.~(\ref{replicatorI}) and~(\ref{resourceScaled}) or Eqs.~(\ref{replicatorII}) and (\ref{resourceScaled}) describes the time-discrete eco-evolutionary dynamics. Henceforth, we term the former system-I and the latter system-II. It, however, must be ascertained that for what values of $S$ and $T$, the type-I and the type-II maps map all initial conditions $x^0\in[0,1]$ to some $x^t\in[0,1]~\forall t$.

\subsection{Choice of Payoff Matrices}
 We fix $R_0=R_1=1$ and $P_0=P_1=0$ to reduce the number of independent parameters in our model. This choice suffices for our purpose as the model still captures all the major ordinal classes of the payoff matrix. A region of the $S$-$T$ parameter space is called a strict physical region~\cite{Pandit2018} for a particular replicator map if the map renders $[0,1]$ interval forward-invariant. It was established~\cite{Pandit2018} that for a payoff matrix, 
\begin{equation}
	\label{Pi}
	{\sf U}=\begin{bmatrix}1 &S\\T &0\end{bmatrix},
\end{equation} 
(where $S$ and $T$ are constants) the strict physical region of the type-I replicator map is a leaf-like region in the $S$-$T$ parameter space, whereas the strict physical region of the type-II map is the non-negative region in the $S$-$T$ parameter space, viz., $S\geqslant0$ and $T\geqslant0$.

As set up earlier, the payoff matrices used in the replicator maps in the eco-evolutionary dynamics have $n^t$-dependence (see Eq.~\ref{effectivePaym}). It can be easily shown that if both ${\sf U}_0$ and ${\sf U}_1$ are chosen such that $(S_0,T_0)$ and $(S_1,T_1)$ are both inside the strict physical region of the $S-T$ parameter space, then the resulting payoff matrix ${\sf U}^t(n^t)$---which is in the form given by Eq.~(\ref{Pi})---has $S$ and $T$ (which now depend on $n^t$) that also remains inside the strict physical region. To see this, we notice that for the type-I maps, the strict physical region is a leaf-like region whose boundary has non-negative unsigned curvature everywhere; and for type-II replicator map the strict physical region is the first octant of the $S-T$ parameter space. Hence, a straight line segment joining of any two points in either of the strict physical regions always remains inside the region; any point on the line is nothing but a convex combination (using factors $n^t$ and $1-n^t$) of the endpoints.

For the purpose of studying TOC, it is ideal to choose ${\sf U}_1$ to represent a game where mutual defection is the dominant strategy. One such game is the Prisoner's Dilemma. Therefore, to write a Prisoner's Dilemma game using the form of $\sf U$ in Eq.~(\ref{Pi}), we choose $S\leqslant0$ and $T\geqslant1$. However, this choice gives us the liberty to choose such payoff matrices for the type-I replicator map that $S$ and $T$ may lie outside the strict physical region of the type-II replicator map. Therefore, we need to use a different class of payoff matrices so that the same payoff matrix can be used in the two eco-evolutionary dynamics---one with type-I and the other with type-II replicator maps; this would bring the results to be obtained for the two replicator maps on same footing so that we can compare and contrast the two dynamics. 

Consider a transformation: ${\sf U}\rightarrow {\sf U}+{\sf 1}$, where $\sf 1$ is a constant matrix with unity as every element. Under this transformation, the type-I map keeps its form invariant, but the type-II map modifies such that now the strict physical region is given by $S\geqslant-1,~T\geqslant-1$. Ergo, 
\begin{equation}
\label{eq:7}
{\sf U} = \begin{bmatrix}2 &S\\T &1
\end{bmatrix},
\end{equation}
which corresponds to the Prisoner's Dilemma matrix if $S\in[0,1]$ and $T>2$, has the possibility of $S$ and $T$ to lie inside the strict physical regions of both maps for all twelve possible distinct games~\cite{}. Thus, henceforth, we choose ${\sf U}_0$ and ${\sf U}_1$ for both the systems such that their forms are given by Eq.~(\ref{eq:7}).

We assume that in the fully replete case, the players interact through the Prisoner's Dilemma game, ${\sf U}_1$, which has `defect' as its dominant strategy. However, as the resource starts to degrade, cooperation may develop. Consequently, we can allow ${\sf U}_0$ to be the payoff matrix for any of the three major classes of games~\cite{C3MB70602H}, viz., (i) the {Harmony game} (like Harmony I), (ii) the {anti-coordination game} (like the Leader game), and (iii) the {coordination game} (like the Stag-Hunt game). In all these three games, cooperation may be maintained as Nash equilibrium. With such choices, the fate of the eco-evolutionary systems is what we eclectically present in the next section.


\section{Results}
We choose the Prisoner's Dilemma game with ${\sf U}_1$ having $(S_1,T_1)=(0.9,2.1)$ without much loss of generality for our purpose of numerical exercises.  Since we wish to investigate the role played by the growth rate $r$ of the resource in averting TOC, the harvesting efforts $e_L$ and $e_H$ are kept fixed. Again, for the concreteness of numerical exercises, we choose $e_L=1$ and $e_H=1.1$ so that the forward invariance condition is adhered to. It implies that $r$ must lie between $r_-\approx1.103$ to $r_+\approx4.897$. Subsequently, we have discovered the rich dynamical behaviour in the  $x$-$n$ phase plane as we take different ${\sf U}_0$'s which do not have defection as the exclusive Nash equilibrium. 

Since the logistic map~\cite{} is known to exhibit periodic and chaotic outcomes, and so does the type-I replicator map ~\cite{Mukhopadhyay2020, Mukhopadhyay2020b}, both system-I and system-II (i.e., the eco-evolutionary dynamics respectively corresponding to the type-I and type-II maps) can be non-convergent. While we perform linear stability of about the fixed points of the systems, the full non-linear dynamical behaviour can be assessed through careful numerics. We plot the bifurcation diagrams along with the largest Lyapunov exponent~\cite{Argyris2015} of the systems and infer the character of TOC from there. Let us now go through some examples illustrating some interesting observations.
\subsection{Illustrative examples}
\label{sec:ie}
\begin{figure}
	\includegraphics[width=1\columnwidth]{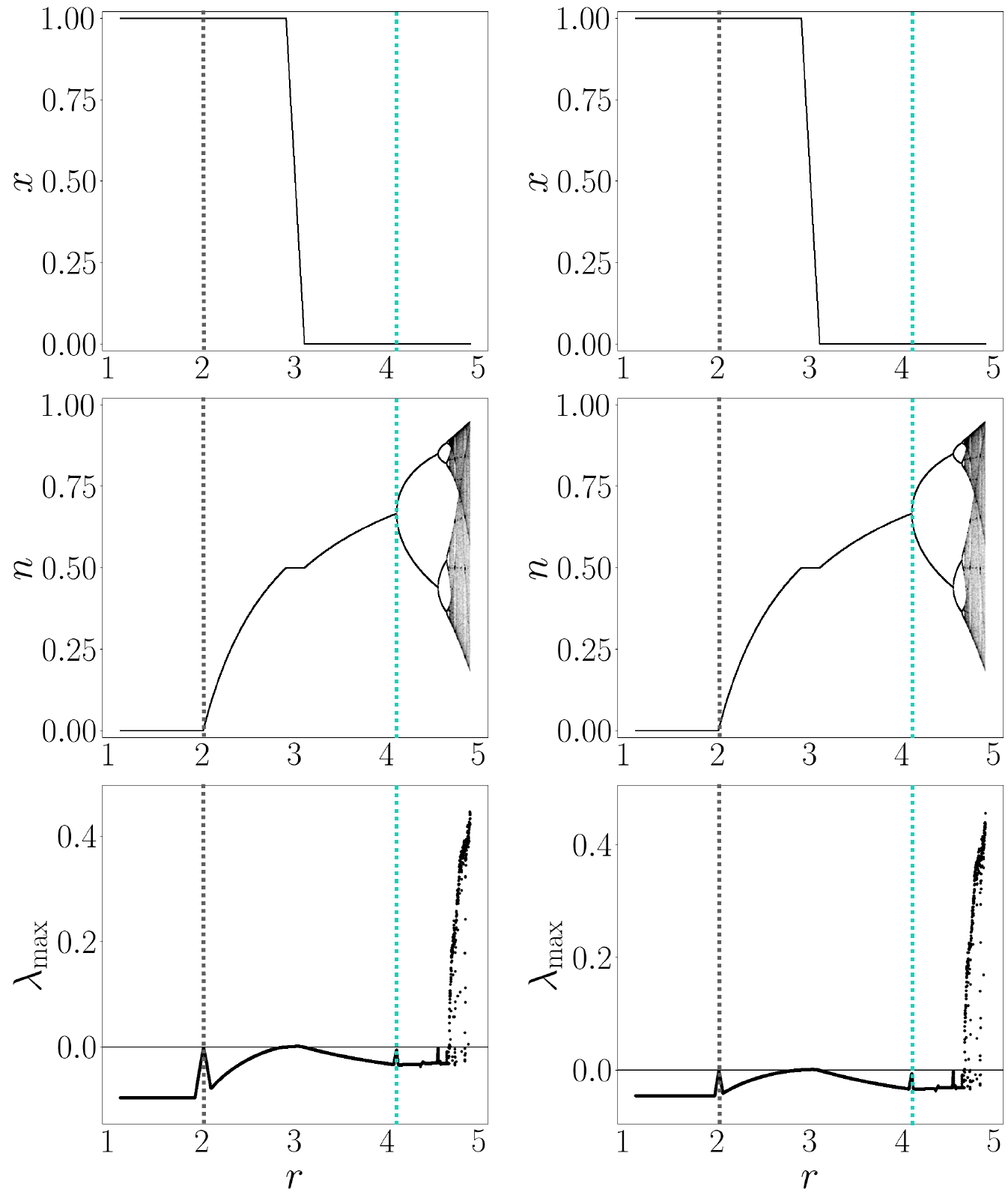}
	\caption{\emph{Prevention of TOC in the resource state which chaotically oscillates, in the complete absence of cooperators, for large values of the intrinsic growth rate:} This figure exhibits bifurcation diagram and maximum Lyapunov exponent for the case with ${\sf U_0}$ having $S_0=1.1$ and $T_0=1.9$, i.e., Harmony-I game. We fix $e_L=1$, $e_H=1.1$, and $r\in\left[1.103,4.897\right]$. The first column is for system-I and the second one is for system-II. The first two rows exhibit the stable solutions (fixed points, periodic orbits, or chaos) resulting from bifurcations as seen respectively in the variables for low harvester fraction ($x$) and (normalized) state of the resource (n). The third row represents the plot of the maximum Lyapunov exponent ($\lambda_{\rm max}$) needed to ascertain the regions of chaos ($\lambda_{\rm max}>0$). The transcritical bifurcation that happens at $r=r_c$ is marked by the grey dashed line and the flip bifurcation leading to period doubling is marked by the blue dotted line. }
	\label{fig:1}
\end{figure}

\textit{Example 1 (Harmony game; $S_0=1.1,~T_0=1.9$---Harmony I game)}: Suppose the harvesters play a Harmony game when the resource is completely depleted. The eco-evolutionary dynamics generates Fig.~\ref{fig:1}, where we note that both systems I and II are almost identical. As the intrinsic growth rate ($r$) of the resource increases from its least possible value ($r_-\approx 1.103$), initially, the resource remains depleted with the consumer population exclusively consisting of low-harvesters. Beyond a critical $r=r_c$ (here, $r_c=2$---the transcritical bifurcation point), the resource starts to achieve non-zero asymptotic values. As $r$ increases more, a flip bifurcation takes place at around $r\approx4$ and the period doubles leading to a stable oscillatory resource state. Finally, chaos appears through the period-doubling route and emerges for many values of $r$---except for some interspersing periodic windows. From the point of flip bifurcation onwards, the harvester population has no low-harvester: The oscillatory (periodic or chaotic) state of the resource is sustained by harvester population consisting of high-harvesters only. Thus, oscillatory outcome helps in preventing TOC even in the absence of any cooperators. A very interesting regime in this example is the flat horizontal plateau around $(r=3,n=0.5)$ in $r$-$n$ plot: One notes that the stable consumer population composition linearly changes from all low-harvesters to all high-harvesters while the resource state remains unchanged. 

\begin{figure}
	\includegraphics[width=1\columnwidth]{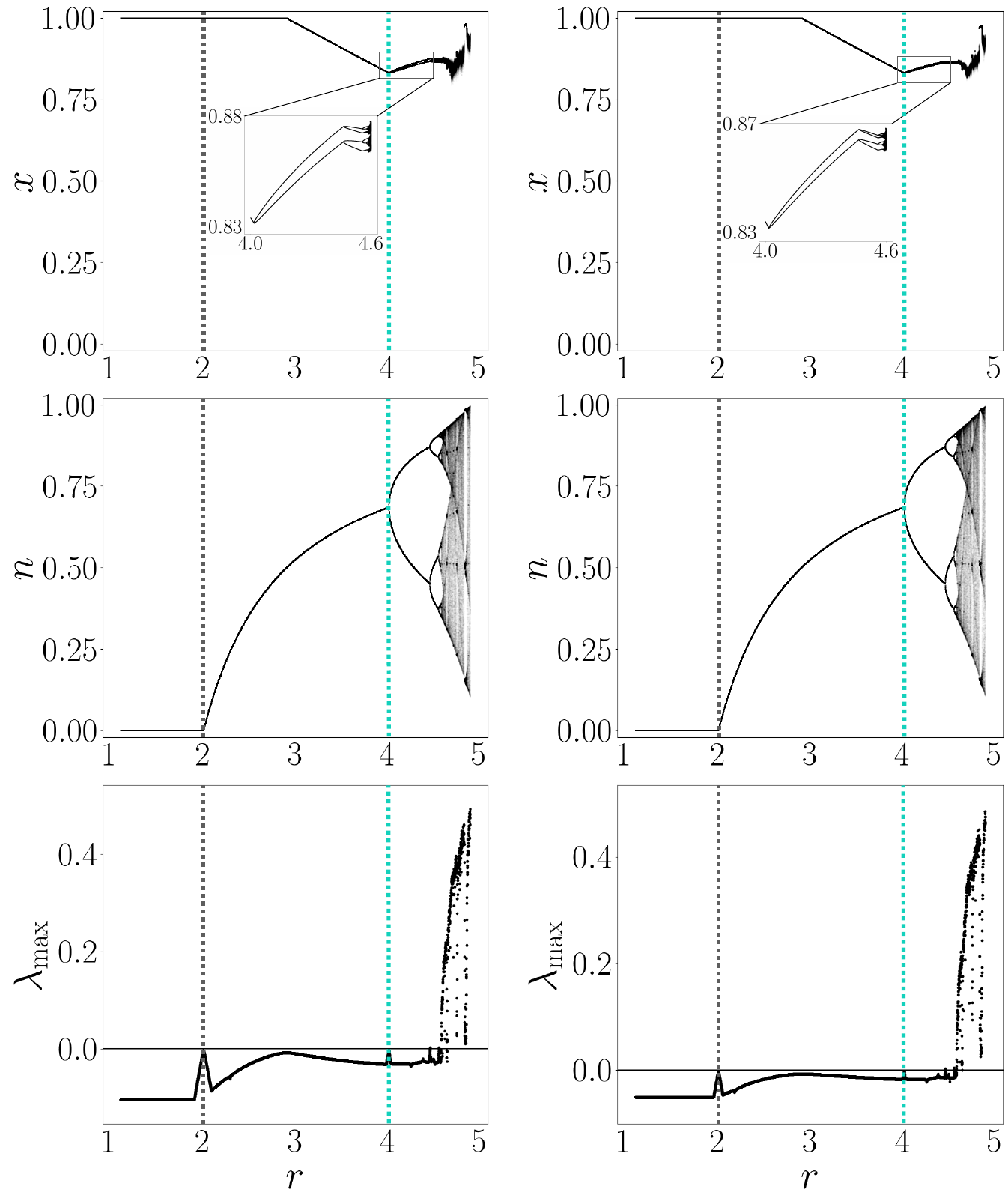}
	\caption{\emph{Prevention of TOC in the resource state which chaotically oscillates in the partial presence of cooperators with chaotically oscillating fraction, for large values of the intrinsic growth rate:} This figure exhibits bifurcation diagram and maximum Lyapunov exponents for the case with ${\sf U_0}$ having $S_0=1.8$ and $T_0=1.9$, i.e., Harmony-I game. Other details are analogous to the ones given in the caption of Fig.~\ref{fig:1}.}
	\label{fig:2}
\end{figure}

\textit{Example 2 (Harmony game; $S_0=1.8,~T_0=1.9$---Harmony I game)}: The aforementioned horizontal flat region, however, is not a generic feature. If we take a slightly different Harmony game $(S_0=1.8,~T_0=1.9)$, we note (see Fig.~\ref{fig:2}) its absence. In this particular game as $r$ increases, the resource state again undergoes period-doubling bifurcation. However, in this case, the low-harvester fraction never vanishes completely; unlike the preceding example, here both $x$ and $n$ coordinates chaotically oscillate in unison. Thus, avoidance of TOC is sustained by chaotically varying fractions of harvester types. These features are the same in both systems I and II.

\begin{figure}
	\includegraphics[width=1\columnwidth]{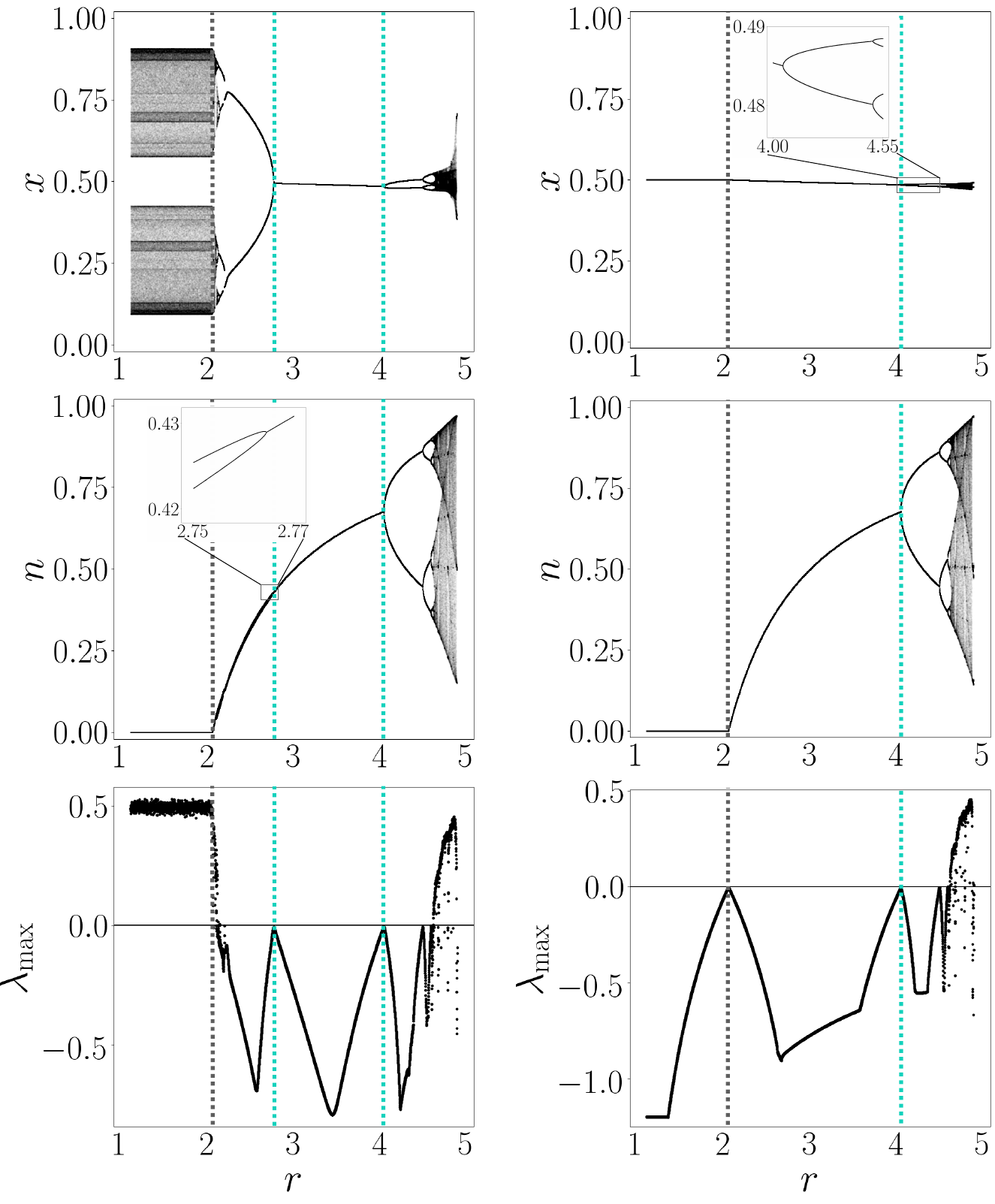}
	\caption{\emph{Prevention of TOC in the resource state which chaotically oscillates in the partial presence of cooperators with chaotically oscillating fraction, for large values of the intrinsic growth rate:} This figure exhibits bifurcation diagram and maximum Lyapunov exponents for the case with ${\sf U_0}$ having $S_0=8$ and $T_0=9$, i.e., Leader game. We fix $e_L=1$, $e_H=1.1$, and $r\in\left[1.103,4.897\right]$. The first and the columns are, respectively, for system-I and system-II.  The stable solutions (fixed points, periodic orbits, or chaos) resulting from bifurcations as seen in the variables for low harvester fraction ($x$) and (normalized) state of the resource (n) in the first and the second row, respectively. The third row showcases the maximum Lyapunov exponent ($\lambda_{\rm max}$) marking the regions of chaos ($\lambda_{\rm max}>0$). The flip bifurcations leading to period doubling are marked by blue dotted lines. The gray dashed line marks the value of $r=r_c$ up to which the TOC is inevitable.}
	\label{fig:3}
\end{figure}

\textit{Example 3 (Coordination game; $S_0=0.9, T_0=1.9$---Stag-Hunt game)}: Next, we take the example of a coordination game. We observe that the features are similar to that of the case of example 1 except that bistability is witnessed. Specifically, when $r$ is just more than $2$, the low-harvesters take over the consumer population and maintain a non-zero, albeit scarce, state of the resource. However, such a state coexists with another stable state $(x=0,n=0)$. As $r$ crosses $2.1$, the latter state also transitions to a non-zero resource state. In fact, there is a regime of $r$-values (lying approximately between $2.1$ and $2.9$; see the blue curve and the black curve immediately below it in Fig.~\ref{fig:3}) where two distinct non-zero resource states---$1+0.9/(1.1-r)$ and $1+1/(1.1-r)$---are respectively sustained by either all low harvesters or all high-harvesters depending on the initial state of the system under consideration.

\begin{figure}
	\includegraphics[width=1\columnwidth]{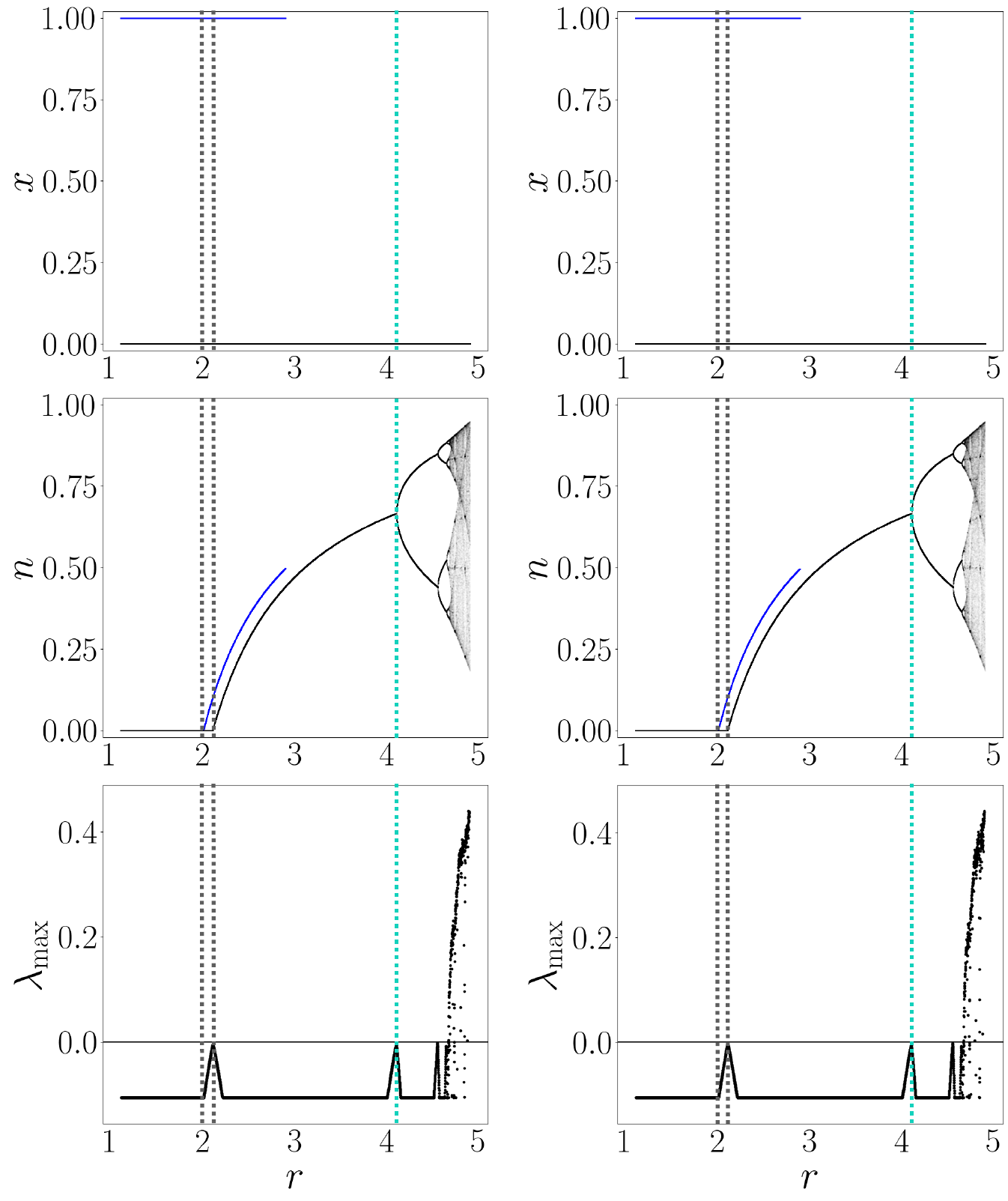}
	\caption{\emph{Prevention of TOC in the resource state which chaotically oscillates, in the complete absence of cooperators, for large values of the intrinsic growth rate:} This figure exhibits bifurcation diagram and maximum Lyapunov exponents for the case with ${\sf U_0}$ having $S_0=0.9$ and $T_0=1.9$, i.e., Stag-Hunt game. Other details are analogous to the ones given in the caption of Fig.~\ref{fig:1}. There are two different values of $r_c$ depending on whether the consumer population is composed of exclusively high harvesters ($x=1$) or low harvesters ($x=0$). The two different coexisting outcomes (fixed points) are marked with two different colours---black and blue. Note that the blue curve and the part of black curve immediately below it indicate bistability: Two distinct resource states, given by $1+0.9/(1.1-r)$ and $1+1/(1.1-r)$, are maintained by exclusively low harvesters and exclusively high-harvesters, respectively.}
	\label{fig:4}
\end{figure}

\textit{Example 4 (Anti-coordination game; $S_0=8, T_0=9$---Leader game)}:
The case of anti-coordination games is further interesting because systems I and II have qualitatively different dynamics for smaller values of intrinsic growth rate. The bifurcation diagrams and maximum Lyapunov exponent are depicted in Fig.~(\ref{fig:4}). This is because the type-I replicator dynamics is chaotic for such Leader games, but type-II is not; at small $n$-values, the dynamics is dominated by $x$ variable. Note that for higher values of $r$ ($\gtrsim3$), the features in the case of the Leader game chosen are the same as that of example 2. Coming to the lower values of $r$, we note that in system II, the non-zero state of the resource is sustained by a mix of low and high harvesters beyond a threshold value of $r$, viz., $r=2.05$; below the threshold, however, the low and high harvesters remain in equal proportions but the TOC is unavoidable. In the case of system I, for $r$ less than $2.05$ begets TOC but now the evolution of harvester-fractions is chaotic. As soon as $r$ crosses $2.05$, both $x$ and $n$ chaotically evolves. With further increase in $r$-value, the chaos is ultimately replaced by periodic oscillations and finally convergent fixed point solutions. In summary, the harvester-fractions are chaotic for both high and low $r$ values, but in the former TOC is averted through chaos in $n$-variable but in the latter TOC is unavoidable even in the presence of low-harvesters.

\begin{figure}[t!]
	\includegraphics[width=0.95\columnwidth]{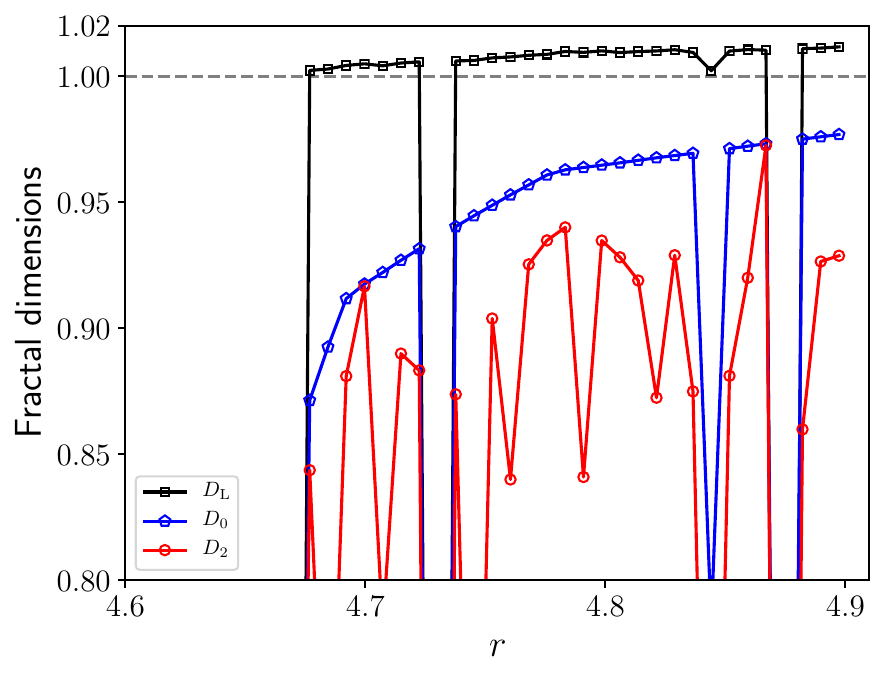}
	\caption{Illustrating hypochaos: Fractal dimension of strange chaotic attractor for ${\sf U_0}$ having $S_0=0.9$ and $T_0=1.9$, i.e., Stag-Hunt game (see Fig.~\ref{fig:2}), in system I. Black squares, blue pentagons, and red circles represent Lyapunov ($D_{\rm L}$), capacity ($D_0$), and correlation ($D_2$) dimensions, respectively, of the attractors at corresponding $r$-values. (Solid lines merely connect the markers.) Note $D_0<1<D_{\rm L}$ signifying hypochaos.}
	\label{fig:DKY}
\end{figure}

In the above examples (specifically in examples 1, 2, and 4), there are instances where chaos is exhibited by either $x$ or  $n$ variables (but not both simultaneously). It is obvious that since the systems are bounded and dissipative, such chaotic regimes in the two-dimensional eco-evolutionary maps must be characterized by Lyapunov spectra with two Lyapunov exponents---one negative ($\lambda_{\rm min}$) and another positive ($\lambda_{\rm max}<\lambda_{\rm min}$). The Lyapunov dimension ($D_{\rm L}$) is, thus, $1+(\lambda_{\rm max}+\lambda_{\rm min})/|\lambda_{\rm min}|>1$. However, the phase points of the corresponding attractor lie on a straight line, and hence the attractor's capacity dimension ($D_0$) cannot be greater than one; in fact, it is less than one (see, e.g., Fig.~\ref{fig:DKY}). Hence, by definition, such instances of chaos can be recognized as hypochaos.

Of course, one could have taken many other possible values of $S_0$ and $T_0$ to check for other possibilities using stability analyses and numerics---which we have exhaustively done in the backdrop---but same features as illustrated by the above four examples would have been found. We have essentially highlighted bistability, $r$-independent resource state (flat plateau in Figs.~\ref{fig:3}), hypochaos, and averting TOC through chaos. Expect for the bistability, the other features are very much particular to the discrete-time dynamics of the generation-wise non-overlapping population; continuous-time eco-evolutionary dynamics~\cite{weitz2016pnas,lin2019prl,tilman2020nc, Mondal2022, Bairagya2021, Bairagya2023} are known not to exhibit these dynamical features. 
\subsection{Comprehending the generic features}
\begin{figure*}[t!]
	\includegraphics[width=1.0\textwidth]{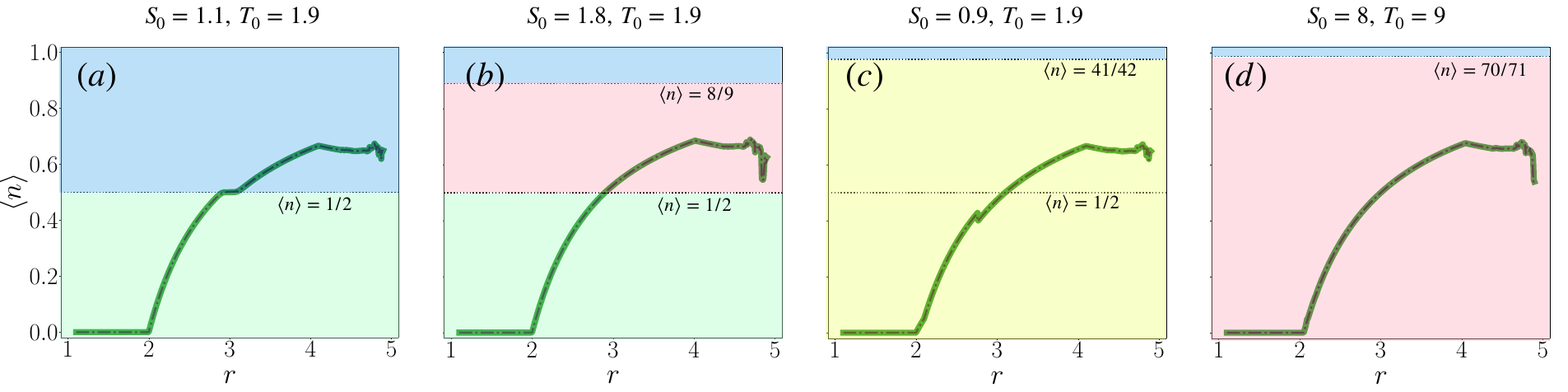}
	\caption{Mean game dictates long-term dynamics: Figure presents $\langle{n}\rangle$ vs. $r$ plot for the four illustrative examples as listed in Sec.~\ref{sec:ie}, viz., ($a$) Example 1: $S_0=1.1$ and $T_0=1.9$, ($b$) Example 2: $S_0=1.8$ and $T_0=1.9$, ($c$) Example 3: $S_0=0.9$ and $T_0=1.9$, and ($d$) Example 4: $S_0=8$ and $T_0=9$. Green solid and black dashed curves correspond to system I and system II, respectively. The background colours---blue, green, pink, and yellow---indicate that the mean game matrix for the corresponding values of $\langle{n}\rangle$ is that of Prisoner's Dilemma game, Harmony game, anti-coordination game, and coordination game.}
	\label{fig:1rn}
\end{figure*}
{\begin{table}[b!]
		\scriptsize
		\begin{center}
\caption{Condition on resource state ($n$) for local asymptotic stability of $x=0$ and $x=1$ manifolds for the case of four illustrative examples (discussed in Sec.~\ref{sec:ie}) with different ${\sf U}_0$. Conclusions are identical for system I and system II.}
			\label{table:1}
			\hspace{1mm}
			\begin{tabular}{c|c|l}
				\hline
				\bf{${\sf U}_0$}                    &\bf{Invariant Manifold}    &\bf{Attractive if}       \\
				\hline
				\hline
				\multirow{2}{*}{Harmony I game}          &\multirow{2}{*}{$x=0$}   	&\\[-5.0pt]
									   			    &                           &$~n\in(\frac{1}{2},1]$   \\[1.5pt]\cline{2-3}
				\multirow{2}{*}{$(S_0=1.1,~T_0=1.9)$} &\multirow{2}{*}{$x=1$} 	&\\[-5.0pt]
				                                    &                           &$~n\in[0,\frac{1}{2})$   \\[1.5pt]\cline{2-3}
				\hline
			    \multirow{2}{*}{Harmony I game}          &\multirow{2}{*}{$x=0$}     &\\[-5.0pt]
				                                    &                           &$~n\in(\frac{8}{9},1]$   \\[1.5pt]\cline{2-3}
				\multirow{2}{*}{$(S_0=1.8,~T_0=1.9)$} &\multirow{2}{*}{$x=1$} 	&\\[-5.0pt]
				                                    &                           &$~n\in[0,\frac{1}{2})$   \\[1.5pt]\cline{2-3}
				\hline
				\multirow{2}{*}{Stag-Hunt game}          &\multirow{2}{*}{$x=0$}     &\\[-5.0pt]
				                                    &                           &$~n\in[0,1]$             \\[1.5pt]\cline{2-3}
				\multirow{2}{*}{$(S_0=0.9,~T_0=1.9)$} &\multirow{2}{*}{$x=1$} 	&\\[-5.0pt]
				                                    &                           &$~n\in[0,\frac{1}{2})$   \\[1.5pt]\cline{2-3}
				\hline
			   \multirow{2}{*}{Leader game}              &\multirow{2}{*}{$x=0$}     &\\[-5.0pt]
				                                    &                           &$~n\in(\frac{70}{71},1]$ \\[1.5pt]\cline{2-3}
			    \multirow{2}{*}{$(S_0=8,~T_0=9)$}     &\multirow{2}{*}{$x=1$} 	&\\[-5.0pt]
				                                    &                           &$~n\in \{~\}$            \\[1.5pt]\cline{2-3}
				\hline
				\hline
			\end{tabular}
		\end{center}
\end{table}}
Let us now try to understand the reasons behind certain generic features, like (i) the existence of a threshold value of $r$ after which $n$ becomes non-zero, (ii) simultaneous chaotic evolution of $x$ and $n$ variables, e.g., in examples 2 and 4, and (iii) hypochaos resulting from chaos exhibited by only one of $x$ and  $n$ variables, e.g., in examples 1, 2 and 4.

It is not surprising that for all choices of ${\sf U}_0$, TOC is found to be inevitable for small values of $r$, while it may be only partially avoided---either through periodic orbits or the chaotic ones---for larger values of $r$. The critical value of resource growth rate $r=r_{c}$ (say), below which state of the resource $n^t$ always fixates to zero, happens to be dependent on the choice of the payoff matrix $U_{0}$ and the type of the replicator map being used. Its value may simply be determined by substituting the asymptotic value of $n^t=n^*=0$ in the replicator equations and finding what values of $x^t=x^*$ are allowed at $r=r_-$; and then by using that particular value of $x^*$ in the effective growth rate, $r-[e_Lx^*+e_H\left(1-x^*\right)]$, of $n^t$ (see Eq.~\ref{eq:8}): The maximum value of $r$ at which the maximum value of effective growth rate remains less than or equal to unity is recognized as $r_c$. In other words, at $(x^*, n^*=0)$ becomes unstable beyond $r_c$ and the TOC is averted. 

For example, when ${\sf U_0}$ corresponds to Harmony game with $S_0=1.1$ and $T_0=1.9$ (Fig.~\ref{fig:1}) or with $S_0=1.8$ and $T_0=1.9$ (Fig.~\ref{fig:2}), one find $x^*=1$ at $r=r_-\approx1.103$. Hence, $r_c=1+[e_Lx^*+e_H\left(1-x^*\right)]=2$. Likewise, when ${\sf U_0}$ corresponds to a Stag-Hunt game with $S_0=0.9$ and $T_0=1.9$ (Fig.~\ref{fig:4}), $r_c$ is either $2$ or $2.1$ depending whether the initial conditions are in the basin of attraction of the fixed point with $x^*=1$ or $x^*=0$, respectively. Of course, the cases where $x^t$ at $r=r_-$ and $n^*=0$ does not converge to a fixed point, e.g., when ${\sf U_0}$ corresponds to Leader game with $S_0=8$ and $T_0=9$ (Fig.~\ref{fig:3}), $r_c$ may be evaluated using $r_c=1+[e_Lx^*+e_H\left(1-x^*\right)]$ where $x^*$ is replaced by the mean value of $x^t$ in the chaotic attractor. For the specific case at hand (see top row of Fig.~\ref{fig:3}), one finds that the mean value is $0.5$ (when type-I map is used) and hence, $r_c=2.05$. The value is the same if type-II map is used, because $x^*=0.5$.

As pointed out earlier, chaos is not always seen in players' fractions even if the coupled dynamics of the resource is chaotic, or the other way round---such a situation at hand is a physical manifestation of hypochaos. For which sets of parameter values, hypochaos appears can be understood mathematically through the stability of invariant manifolds and physically through the idea of \textit{mean game}. The mean game is defined as the game with payoff matrix, $\langle{\sf U}\rangle \equiv (1-\langle{n}\rangle) {\sf U_0} + \langle{n}\rangle {\sf U_1}$, where $\langle{n}\rangle$ is the mean value of $n^t$ over the time-series of $n^t$ on the attractor of corresponding system. For example, when ${\sf U_0}$ corresponds to a Harmony game with $S_0=1.1$ and $T_0=1.9$, $n^t$ varies chaotically at $r=4.8$; for various $n^t$ values the effective payoff matrix  ${\sf U}^t(n^t) = (1-n^t) {\sf U_0} + n^t {\sf U_1}$ fluctuates between that for Harmony game and Prisoner's Dilemma. The mean payoff matrix, $\langle{\sf U}\rangle$, however, can be shown to be that of the Prisoner's Dilemma. Naturally, the cooperator fraction should vanish in such a scenario (as seen in Fig.~\ref{fig:1}a and Fig.~\ref{fig:1}d).

There are two invariant manifolds, $x=0$ and $x=1$, as far as Eq.~(\ref{replicatorI}) and Eq.~(\ref{replicatorII}) are concerned. One can find their linear stabilities (along $x$-direction) by calculating $\partial x^{t+1}/\partial x^t$ at each of the manifolds and by finding if its absolute value is less than unity at the corresponding manifold for which values of $n$. The conclusions of the stability analysis are summarized in Table~\ref{table:1}. If $n$ varies chaotically (or even periodically) in the asymptotic limit, then we propose the ansatz that the stability of the invariant manifolds can be ascertained by using $\langle n\rangle$ in place of $n^t$ in $\partial x^{t+1}/\partial x^t$. This ansatz about replacing a fluctuating payoff matrix in the replicator equations with the mean game payoff matrix implicitly assumes that the distribution of the phase points on the attractor is uniform. 

To understand the mathematical reason behind the hypochaos, let us focus as an example on Fig.~\ref{fig:1} and Fig.~\ref{fig:1rn}a (and Table~\ref{table:1}), i.e., example 1 of Sec.~\ref{sec:ie}. We note that the invariant manifolds are stable for mutually exclusive ranges of $n$. For values of $r$ when $n$ is no longer asymptotically convergent, which manifold is stable can be predicted by finding $\langle n\rangle$ for these cases. It is seen to lie between $1/2$ to $1$ for both systems I and II (see green solid and black dot-dashed curves, respectively, in Fig.~\ref{fig:1rn}a). Obviously, the $x=0$ manifold is stable in such case making the chaos a hypochaotic one. Moreover, in this (blue-coloured) region of $n$-values, the mean game payoff matrix corresponds to Prisoner's Dilemma. Hence, it is not surprising that defectors (i.e., high harvesters) fixate in the consumer population while the resource varies periodically or chaotically. All other instances of hypochaos (examples 1, 2, and 4) can be understood similarly.

The case of both variables behaving chaotically say in example 2 (Fig.~\ref{fig:2} and Fig.~\ref{fig:1rn}b) for $r\gtrsim4$, can be reasoned along the same line. We note that after period-doubling starts,  $1/2<\langle n\rangle<8/9$. In this region neither of the invariant manifolds is stable (see Table~\ref{table:1}); the region corresponds to a mean game payoff matrix of anti-coordination game making it reasonable the mixed states of the population is sustained. Obviously, since there is no other invariant manifold (except irrelevant to this case: $n=0$), oscillatory $n$ leads to oscillatory $x$, and generically, the attractor dimension is more than unity; i.e., hypochaos is not present in this case. 

We end this section by making two brief remarks. First, as far as non-convergent $n$ is concerned, the features of example 2 (Fig.~\ref{fig:3} and Fig.~\ref{fig:1rn}c) are very similar to the aforementioned discussion of example 1. The only difference is that rather than the Prisoner's Dilemma game, the mean game corresponds to the coordination game (yellow region) which has two pure and one mixed state equilibria. In this case, however, only $x=0$ is stable (see Table~\ref{table:1}) when $1/2<\langle n\rangle<41/42$; hence, high harvesters fixate in the consumer population sustaining a chaotically fluctuating resource population. Second, just as we use $\langle n\rangle$, one can use $\langle x\rangle$ in example 4 (Fig.~\ref{fig:4} and Fig.~\ref{fig:1rn}d) for $r<r_c$ to find that $n=0$ manifold is stable for $x=\langle x\rangle=0.5$ and hence, appears an instance of hypochaos with attractor lying on $n=0$.

\section{Discussion and Conclusions}
\label{sec:conclusion}

Some investigations~\cite{weitz2016pnas,lin2019prl,Bairagya2021,Mondal2022,Liu2023} have explored scenarios close to the one examined herein. Specifically, these scenarios are where resources for individual consumption are generated by cooperators within an infinitely large unstructured population---the kind of simple population structure that we are interested in this paper. The interplay of types composing the population influences the dynamics of the shared resource, and feedback from the shared resource state impacts the interaction among the types.

Somewhat along similar line, some papers~\cite{Chen2018,tilman2020nc,Yan2021} have modelled an infinitely large unstructured consumer population engaged in the consumption of a self-renewable resource---one of the main ingredients of this paper. However, the intrinsic growth rate in those papers carries a different implication compared to ours. In those studies, where the resource dynamics are governed by a continuous-time dynamics, the positivity of the effective intrinsic growth rate ensures the resource state's  growth. In contrast, in our study, we have aimed to model a dynamic resource with non-overlapping generations, and hence, the growth of the resource state is ensured only when the (effective) intrinsic growth rate at any given time exceeds unity. Also, due to the restriction of forward invariance of the phase space, our discrete-time model needs the intrinsic growth rate to be always more than the harvesting rates; this is not a necessary requirement in continuous time models~\cite{Bairagya2023}.

Furthermore, recent researches~\cite{Chen2018,Mondal2022} have also delved into inquiries exploring the impacts of external motivators, such as incentives for resource preservation, penalties for damage to it, and inspection imposed on the consumer population. Another paper~\cite{lin2019prl} has also included the spatial structure of the consumer population. Another notable extension~\cite{Yan2021} of the aforementioned models considers a time-lagged effect of the consumer population on the dynamics of a self-renewing resource. All these additional dimensions of inquiry have remained beyond the scope of our present study. Also,  variations~\cite{Bairagya2021,Bairagya2023}, within these classes of model, explore scenarios where a consumer population is growing to reach its carrying capacity---something we ignore in the present study.

The deterministic dynamics that we have dealt with in this paper can actually be seen as a mean-field model of a microscopic stochastic birth-death process in the populations. Some previous studies~\cite{lin2019prl,Bairagya2023} have aimed to elucidate the dynamics of the consumer population and the shared resource from a microscopic point of view. These investigations unveil that the effective deterministic dynamics, dictating the evolutions of types' frequencies within the population, population size, and the state of the shared resource, can be deduced as mean-field dynamics from more fundamental microscopic descriptions of the system. However, in this paper, we have not  derived the dynamical equations of difference equations from any microscopic description; rather, the deterministic equations have been formulated phenomenologically.

Moreover, one can actually, in principle, go beyond the mean-field description and investigate the full underlying stochastic equations governing the system. In this context, it is worth pointing out some works~\cite{Huang2015,Wienand2017,stollmeier2018unfair} that delved into the impact of stochasticity on eco-evolutionary systems of this nature. However, these investigations presuppose that environmental fluctuations arise independently of evolutionary individuals while attributing the fluctuations to intrinsic factors. Consequently, these studies analyze the repercussions of fluctuations that are explicitly unrelated to the consumer population and disregard the feedback to the environment influenced by the changing fractions of types of individuals within the population. It is definitely of interest to extend such studies to the exact scenario considered in this paper.

In the light of the above discussion, in this paper, our goal may initially appear rather modest in the sense that we have extended the deterministic eco-evolutionary dynamics \`a la Tilman et al.~\cite{tilman2020nc} to the unexplored case of generation-wise non-overlapping populations of consumers harvesting self-renewing resource. Nevertheless, the dynamics is much more rich with the appearance of chaos---something totally absent in the corresponding continuous case. Moreover, we discover the interesting physical implications of hypochaos: The signature of chaos may be confined to either the dynamics of the resource or that of the consumer fractions. A specific intriguing implication is that resource can keep chaotically evolving, and hence does not vanish to manifest TOC, and yet the consumer population may be composed exclusively of defectors, i.e., high harvesters. This is at odds with the general intuition~\cite{weitz2016pnas} that the prevention of TOC requires cooperation in society to play a decisive role.

We have been rather exhaustive in our choice of dynamical equations: We have used both the type-I and the type-II replicator maps. Additionally, we have used resource-dependent payoff matrices to ensure game-environment feedback. We have ensured forward invariance of the unit square, $[0,1]^2\subset\mathbb{R}^2$ under the dynamics, which leads to appropriate non-trivial bounds on the system parameters. Even though the resource-dependent payoff matrix fluctuates (sometimes chaotically), we have established that the observations regarding the long-time dynamics of the system could be satisfactorily understood through the idea of a mean game payoff matrix.

The stark difference in the fate of the commons in generation-wise overlapping population to that in non-overlapping population should carry over to the finite population as well. Thus, whether one considers the Wright--Fisher process~\cite{Fisher1930,Wright1931} or the Moran process~\cite{Moran1958} for the microevolution of the players in a finite population of consumers harvesting a finite population of self-renewing resource should make a difference in the resulting stochastic eco-evolutionary outcomes. This is what we would like to investigate in the future.

From another perspective, the present study may be seen as a proposal for the simplest deterministic framework of game-resource feedback dynamics where the fate of resource is governed by the chaotic eco-evolutionary dynamics. Of course, chaos can appear even in continuous-time models but the corresponding phase space has to be of higher dimensions and hence, they mostly would be analytically intractable. Nevertheless, how much of the features---e.g., hypochaos-mediated prevention of TOC---presented in the model of this paper carries over to chaotic time-continuous model is also of a possible future direction of research.

\section*{ Acknowledgments} 
SSM is thankful to Vikash Kumar Dubey for help with numerics. SC acknowledges the support from SERB (DST, govt. of India) through project no. MTR/2021/000119.
\section*{AIP Publishing data sharing policy} 
Data sharing not applicable --- no new data generated.
\bibliography{mondal_etal_manuscript}
\end{document}